\documentclass[baaa]{baaa}

\usepackage[pdftex]{hyperref}
\usepackage{subfigure}
\usepackage{natbib}
\usepackage{helvet,soul}
\usepackage[font=small]{caption}

\begin{document}

%%%%%%%%%%%%%%%%%%%%%%%%%%%%%%%%%%%%%%%%%%%%%%%%%%%%%%%%%%%%%%%%%%%%%%%%%%%%%%
%                                                                            %
% Datos de la publicación, no deben ser cambiados.                           %
%                                                                            %
% Journal data, please do not change them.                                   %
%                                                                            %
%%%%%%%%%%%%%%%%%%%%%%%%%%%%%%%%%%%%%%%%%%%%%%%%%%%%%%%%%%%%%%%%%%%%%%%%%%%%%%

\journalvol{61A}
\journalyear{2019}
\journaleditors{R. Gamen, N. Padilla, C. Parisi, F. Iglesias \& M. Sgr\'o}

%%%%%%%%%%%%%%%%%%%%%%%%%%%%%%%%%%%%%%%%%%%%%%%%%%%%%%%%%%%%%%%%%%%%%%%%%%%%%%
%                                                                            %
%  Seleccione el idioma de su contribución: Recuerde que todos los           %
%  componentes del documento (titulo, texto, figuras, tablas, etc.)          %
%  deben estar en el mismo idioma.                                           %
%                                                                            %
%  Select the languague of your contribution: Please remember that all       %
%  document parts (title, text, figures, tables, etc.) must be in the        %
%  same languaje.                                                            %
%                                                                            %
%  0: Castellano / Spanish                                                   %
%  1: Inglés / English                                                       %
%                                                                            %
%%%%%%%%%%%%%%%%%%%%%%%%%%%%%%%%%%%%%%%%%%%%%%%%%%%%%%%%%%%%%%%%%%%%%%%%%%%%%%

\contriblanguage{1}

%%%%%%%%%%%%%%%%%%%%%%%%%%%%%%%%%%%%%%%%%%%%%%%%%%%%%%%%%%%%%%%%%%%%%%%%%%%%%%
%                                                                            %
%  Seleccione el tipo de contribución solicitada:                            %
%                                                                            %
%  Select the requested contribution type:                                   %
%                                                                            %
%  1: Presentación mural / Poster                                            %
%  2: Presentación oral / Oral contribution                                  %
%  3: Informe invitado / Invited report                                      %
%  4: Mesa redonda / Round table                                             %
%  5: Presentación Premio Varsavsky / Varsavsky Prize contribution           %
%  6: Presentación Premio Sahade / Sahade Prize contribution                 %
%  7: Presentación Premio Sérsic / Sérsic Prize contribution                 %
%                                                                            %
%%%%%%%%%%%%%%%%%%%%%%%%%%%%%%%%%%%%%%%%%%%%%%%%%%%%%%%%%%%%%%%%%%%%%%%%%%%%%%

\contribtype{2}

\thematicarea{3}

\title{Star formation at high redshift}
%\subtitle{Instrucciones de estilo}

%%%%%%%%%%%%%%%%%%%%%%%%%%%%%%%%%%%%%%%%%%%%%%%%%%%%%%%%%%%%%%%%%%%%%%%%%%%%%%
%                                                                            %
%  Agregue un título corto para el encabezado de las páginas pares.          %
%                                                                            %
%  Add a short title to appear in the header of even pages.                  %
%                                                                            %
%%%%%%%%%%%%%%%%%%%%%%%%%%%%%%%%%%%%%%%%%%%%%%%%%%%%%%%%%%%%%%%%%%%%%%%%%%%%%%

\titlerunning{Star formation at high redshift}

%%%%%%%%%%%%%%%%%%%%%%%%%%%%%%%%%%%%%%%%%%%%%%%%%%%%%%%%%%%%%%%%%%%%%%%%%%%%%%
%                                                                            %
%  Lista de autores. Los nombres de los autores deben estar separados por    %
%  comas, y deben tener el formato A.E. Autor (iniciales apellido(s);   sin  %
%  coma entre apellido e iniciales ni espacios entre las iniciales).         %
%                                                                            %
%  Author list. Authors' names must be separated by commas, and stick to     %
%  the format A.E. Author (initials Family name -neither commas between      %
%  name and the initials nor blanks between the initials).                   %
%                                                                            %
%%%%%%%%%%%%%%%%%%%%%%%%%%%%%%%%%%%%%%%%%%%%%%%%%%%%%%%%%%%%%%%%%%%%%%%%%%%%%%

\author{FFibla\inst{1}, S. Bovino\inst{1}, R. Riaz\inst{1},  V. B. D\'iaz\inst{1}, C. Olave\inst{1}, S. Vanaverbeke\inst{2}, \& D. R. G. Schleicher\inst{1}}
\authorrunning{FFibla et al.}

%%%%%%%%%%%%%%%%%%%%%%%%%%%%%%%%%%%%%%%%%%%%%%%%%%%%%%%%%%%%%%%%%%%%%%%%%%%%%%
%                                                                            %
% Por favor provea una dirección de e-mail de contacto para los lectores.    %
%                                                                            %
% Please provide a contact e-mail address for the readers.                   %
%                                                                            %
%%%%%%%%%%%%%%%%%%%%%%%%%%%%%%%%%%%%%%%%%%%%%%%%%%%%%%%%%%%%%%%%%%%%%%%%%%%%%%

\contact{pfibla@udec.cl}

\institute{
Departamento de Astronom\'ia, Universidad de Concepci\'on, Barrio Universitario, Concepci\'on, Chile \and Centre for mathematical Plasma-Astrophysics, Department of Mathematics, KU Leuven, Celestijnenlaan 200B, B-3001 Heverlee, Belgium
}

%%%%%%%%%%%%%%%%%%%%%%%%%%%%%%%%%%%%%%%%%%%%%%%%%%%%%%%%%%%%%%%%%%%%%%%%%%%%%%
%                                                                            %
%  El resumen y el abstract son ambos obligatorios, independientemente del   %
%  lenguaje elegido.                                                         %
%                                                                            %
%  The Resumen and the abstract are both mandatory, regardless of the chosen %
%  language.                                                                 %
%                                                                            %
%%%%%%%%%%%%%%%%%%%%%%%%%%%%%%%%%%%%%%%%%%%%%%%%%%%%%%%%%%%%%%%%%%%%%%%%%%%%%%

\resumen{
La importancia del uso de detallados modelos qu\'imicos para comprender la formaci\'on estelar a baja metalicidad ha sido ampliamente reconocida, como ha sido remarcado en recientes investigacinones.

Presentamos aqu\'i simulaciones tridimensionales hidrodin\'amicas para formaci\'on estelar. Nuestro objetivo es explorar el efecto del enfriamiento de la l\'inea de metal sobre la termodin\'amica del proceso de formaci\'on estelar. Exploramos el efecto de cambiar la metalicidad del gas desde $Z/Z_{\odot}=10^{-4}$ hasta $Z/Z_{\odot}=10^{-2}$. Además, exploramos las implicancias de utilizar el patrón de abundancia observacional de una estrella CEMP-no, las cuales han sido propuestas como estrellas de segunda-generaci\'on, la llamada Poblaci\'on III.2.

Para lograr nuestro objetivo, modelamos la micro-f\'isica utilizando el paquete p\'ublico de astroqu\'imica KROME, usando una red qu\'imica que incluye diecis\'eis especies qu\'imicas (H {\sc i}, H {\sc ii}, H$^{-}$, He {\sc i}, He {\sc ii}, He {\sc iii}, e$^{-}$, H$_{2}$ {\sc i}, H$_{2}$ {\sc ii}, C {\sc i}, C {\sc ii}, O {\sc i}, O {\sc ii}, Si {\sc i}, Si {\sc ii}, and Si {\sc iii}). Juntamos KROME con el c\'odigo basado en \textit{Smoothed-particle hydrodynamics} (SPH) tridimiensional hidrodin\'amico GRADSPH. En este contexto de trabajo investigamos el colapso de una nube enriquicida en metales, explorando el proceso de fragmentaci\'on y formaci\'on estelar.

Encontramos que la metalicidad tiene un claro impacto en la termodin\'amica del colapso, permitiendo que la nube alcance la temperatura de piso del CMB a una metalicidad $Z/Z_{\odot}=10^{-2}$, la cual concuerda con trabajos anteriores. Adem\'as, encontramos que al utilizar el patrón de abundancia de la estrella SMSS J031300.36-670839.3 el comportamiento termodin\'amico de la nube es bastante similar a aquellas simulaciones con metalicidad $Z/Z_{\odot} = 10^{-2}$, debido a la alta presencia de carbón. Mientras solo el enfriamiento de la l\'inea de metal sea considerado, los resultados obtenidos confirman el l\'imite de metalicidad propuesto en trabajos anteriores, el cual es muy probable regule los primeros episodios de fragmentaci\'on y potenciamente determine las masas de los conjuntos de estrellas resultantes. Para un completo modelamiento del IMF y su evolución, notamos tambi\'en que el enfr\'iamento producido por el polvo necesita ser considerado.

}

\abstract{
The importance of detailed chemical models to understand low-metallicity star formation is widely recognized, as reflected also in recent investigations. We present here a three-dimesional hydrodynamical simulation for star formation. Our aim is to explore the effect of the metal-line cooling on the thermodynamics of the star-formation process. We explore the effect of changing the metallicty of the gas from $Z/Z_{\odot}=10^{-4}$ to $Z/Z_{\odot}=10^{-2}$. Furthermore, we explore the implications of using the observational abundance pattern of a CEMP-no star, which have been considered to be the missing second-generation stars, the so-called Pop. III.2 stars.

In order to pursue our aim, we modelled the microphysics by employing the public astrochemistry package KROME, using a chemical network which includes sixteen chemical species (H {\sc i}, H {\sc ii}, H$^{-}$, He {\sc i}, He {\sc ii}, He {\sc iii}, e$^{-}$, H$_{2}$ {\sc i}, H$_{2}$ {\sc ii}, C {\sc i}, C {\sc ii}, O {\sc i}, O {\sc ii}, Si {\sc i}, Si {\sc ii}, and Si {\sc iii}). We couple KROME with the fully three-dimensional Smoothed-particle hydrodynamics (SPH) code GRADSPH. With this framework we investigate the collapse of a metal-enhanced cloud, exploring the fragmentation process and the formation of stars.

We found that the metallicity has a clear impact on the thermodynamics of the collapse, allowing the cloud to reach the CMB temperature floor for a metallicity $Z/Z_{\odot}=10^{-2}$, which is in agreement with previous work. Moreover, we found that adopting the abundance pattern given by the star SMSS J031300.36-670839.3 the thermodynamics behavior is very similar to simulations with a metallicity of $Z/Z_{\odot}=10^{-2}$, due to the high carbon abundance. As long as only metal line cooling is considered, our results support the metallicity threshold proposed by previous works, which will very likely regulate the first episode of fragmentation and potentially determine the masses of the resulting star clusters. For a complete modeling of the IMF and its evolution, we expect that also dust cooling needs to be taken into account.
}

%%%%%%%%%%%%%%%%%%%%%%%%%%%%%%%%%%%%%%%%%%%%%%%%%%%%%%%%%%%%%%%%%%%%%%%%%%%%%%
%                                                                            %
%  Seleccione las palabras clave que describen su contribución. Las mismas   %
%  son obligatorias, y deben tomarse de la lista de la American Astronomical %
%  Society (AAS), que se encuentra en la página web indicada abajo.          %
%                                                                            %
%  Select the keywords that describe your contribution. They are mandatory,  %
%  and must be taken from the list of the American Astronomical Society      %
%  (AAS), which is available at the webpage quoted below.                    %
%                                                                            %
%  https://aas.org/authors/astronomical-subject-keywords-update-august-2013  %
%                                                                            %
%%%%%%%%%%%%%%%%%%%%%%%%%%%%%%%%%%%%%%%%%%%%%%%%%%%%%%%%%%%%%%%%%%%%%%%%%%%%%%


\maketitle

\section{Introduction}
\label{ch1}
The birth of the very first stars in the Universe must have occured at redshifts $z \sim 15 - 30$ in dark matter mini-halos with $\sim 10^{6}$ solar masses (e.g., \cite{Haiman1996}, \cite{Tegmark1997}). Such mini-halos were composed from a primordial gas of a few chemical species where the main coolant was molecular hydrogen (\cite{Galli1998}). Thus, the expected temperature for these primordial clouds is  about 300 K, which is thirty times greater than the temperature in typical present-day clouds. The Jeans mass associated with these clouds is therefore greater when compared to their present-day counterparts, as well as the mass of the collapsing objects.\\

From hydrodynamical simulations, it has been shown that several stars could be born from a single dark matter mini-halo, contrary to past results that pointed towards the formation of a single star per dark matter mini-halo. Moreover, the clear impact of some chemical species in the star formation process has been established. In particular, \cite{Bromm2001} proposed that the presence of metals triggers fragmentation in metal deficient primordial clouds up to a metallicity $Z/Z_{o} = 10^{-3.5}$, which has been confirmed by recent investigations, e.g, \cite{Bovino2014}, \cite{Safranek2014}. But, although the qualitative picture of the formation of the Pop III.1 stars is rather well known, how the star formation mode shifts from extremely massive stars with 100-1000 solar masses to present-day stars is still unclear. The development of surveys searching for the most metal-deficient stars has shown that the ratio between oxygen, carbon and nitrogen is enhanced compared to iron for around one-quarter of all the known stars with $\mathrm{[Fe/H]} < -2.0$ (\cite{Beers2005}). These particular stars are now collectively known as carbon-enhanced metal poor (CEMP) stars, and arbitrarily have been defined to have  $\mathrm{[C/Fe]} > +0.7$. Furthermore, subsequent studies have grouped the stars falling into the CEMP-definition into four different sub-groups on the basis of the abundances of their electron-capture associated species. The CEMP-s stars show an overabundance of chemical species produced by the s-process, the CEMP-r stars show an overabundance of chemical species produced by the r-process, the CEMP-s/r stars show an overabundance of elements related to both processes, while the CEMP-no stars do not show an overabundance of elements, neither related to the s-process or related to the r-process.\\

Abundances for the s-group are well explained by means of mass transfer in a binary system from an AGB star to a secondary smaller star which is the one observed today. For the CEMP-no group the panorama is a bit more complicated as several progenitors have been proposed by different authors. As has been found by \cite{Hansen2016b}, CEMP-no stars seem to be bona-fide second-generation stars. This has been proposed on the basis of multiple observational findings (e.g., \cite{Cook2011}, \cite{Cook2012}), \cite{Carollo2012},   \cite{Yoon2016}, \cite{Hansen2016b}).\\

The paper is structured as follows. Sec. \ref{ch2} describes the computational scheme employed in our simulations, as well as the initial conditions for all our models and its features. Sec. \ref{ch3} contains a description from our results, while our conclusions are presented in Sec. \ref{ch4}

\section{Methods}
\label{ch2}
Simulations were carried out by combining two different codes. One was GRADSPH\footnote{\url{https://www.swmath.org/software/1046}} , developed by \cite{Vanaverbeke2009}, a parallel fully three-dimensional TREESPH code designed to evolve self-gravitating astrophysical fluids. The other one was KROME\footnote{\url{http://www.kromepackage.org}}, developed by \cite{Grassi2014}, a novel astrochemical open-source package to treat the microphysics of the collapse, such as the temperature and the evolution of the chemical species included networks used. Such framework has been already used by \cite{Riaz2018c}, investigating primordial star formation and its binaries properties. GRADSPH has been further tested on star formation and their evolution in \cite{Riaz2018a} and \cite{Riaz2018b}.\\

We perform several simulations, varying the initial metallicity of the cloud from a primordial case to  $Z/Z_{\odot}=10^{-2}$, including the one given by the observational pattern of the Keller star \citep{Keller2014}. All simulations were started at the same redshift $z = 15.0$, an initial temperature $\mathrm{T} = 300$ K, and an initial density $\rho = 10^{-22}$ g $\cdot$ cm$^{-3}$, from which in order to assure the gravitational collapse of the clouds, the initial mass was set as $\mathrm{M}_{J} = 1.026\times10^{6}$ $\mathrm{M}_{\odot}$. Moreover, we defined two groups of simulations based on their chemical pattern, labeled as p-runs, for the one using a primordial network which includes nine chemical species: H {\sc i}, H {\sc ii}, He {\sc i}, He {\sc ii}, He {\sc iii}, e$^{-}$, H$_{2}$ {\sc i}, H$_{2}$ {\sc ii}, and H$^{-}$; and the m-runs, for the ones using a metal-enriched network which includes the already named species for the p-run, plus the metal-species C {\sc i}, C {\sc ii}, O {\sc i}, O {\sc ii}, Si {\sc i}, Si {\sc ii}, and Si {\sc iii}. All species were initialized in number densities, with a value of almost zero ($n_{X} = 10^{-40}$ cm$^{-3}$), with the exception of H {\sc i}, He {\sc i}, H$_{2}$ {\sc i}, H {\sc ii}, e$^{-}$, C {\sc ii} (carbon was assumed as totally ionized), O {\sc i}, and Si {\sc i}. The non-metal species were initialized in number denisities as $n_{\mathrm{H}} = 44.81$, $n_{\mathrm{He}} = 3.72$, $n_{\mathrm{H}_{2}} = 2.98\times10^{-5}$, and $n_{\mathrm{H}^{+}} = 5.97\times10^{-3}$. The metal species were computed on-the-fly by KROME for models met2 ($Z/Z_{\odot}=10^{-2}$) and met3 ($Z/Z_{\odot}=10^{-3}$) and scaled according to their metallicities, in the Keller model the reported observed abundances (\cite{Keller2014}) were used. The initial abundance of the electrons was computed on-the-fly by KROME for all models, such that the positive charge of the species was balanced.

\section{Results}
\label{ch3}
\begin{figure}[!t]
\centering
\includegraphics[width=0.45\textwidth]{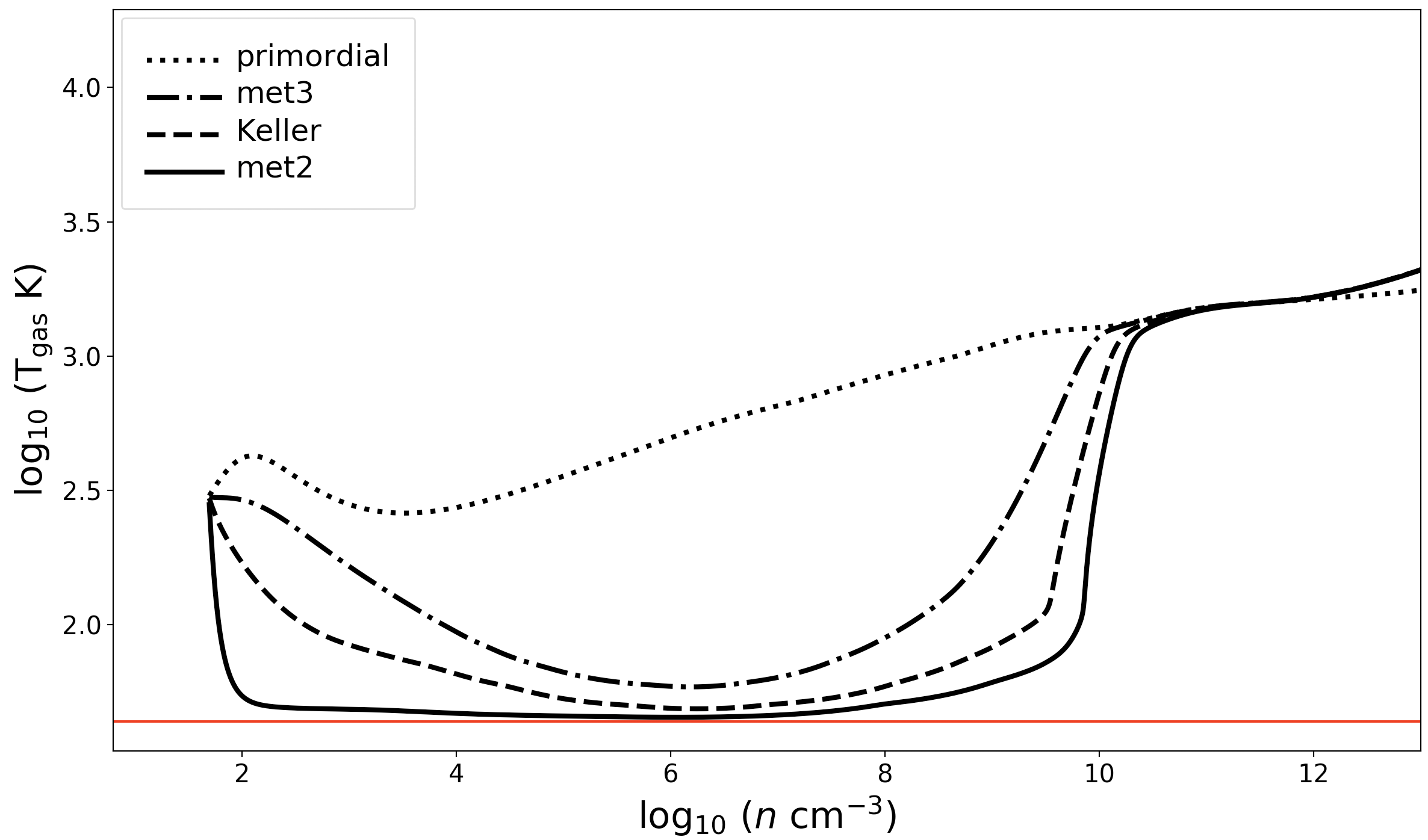}
\caption{Density profile of the temperature evolution of prestellar clouds with different chemical species abundances. The dotted line represents the primordial model, the dotted-dashed line the met3 model, the dashed line the met2 model, and the solid line the Keller model. The red bottom line represents the CMB floor temperature give by the initial redshift of the simulations.}
\label{n_T}
\end{figure}

\begin{figure}[!t]
\centering
\includegraphics[width=0.45\textwidth]{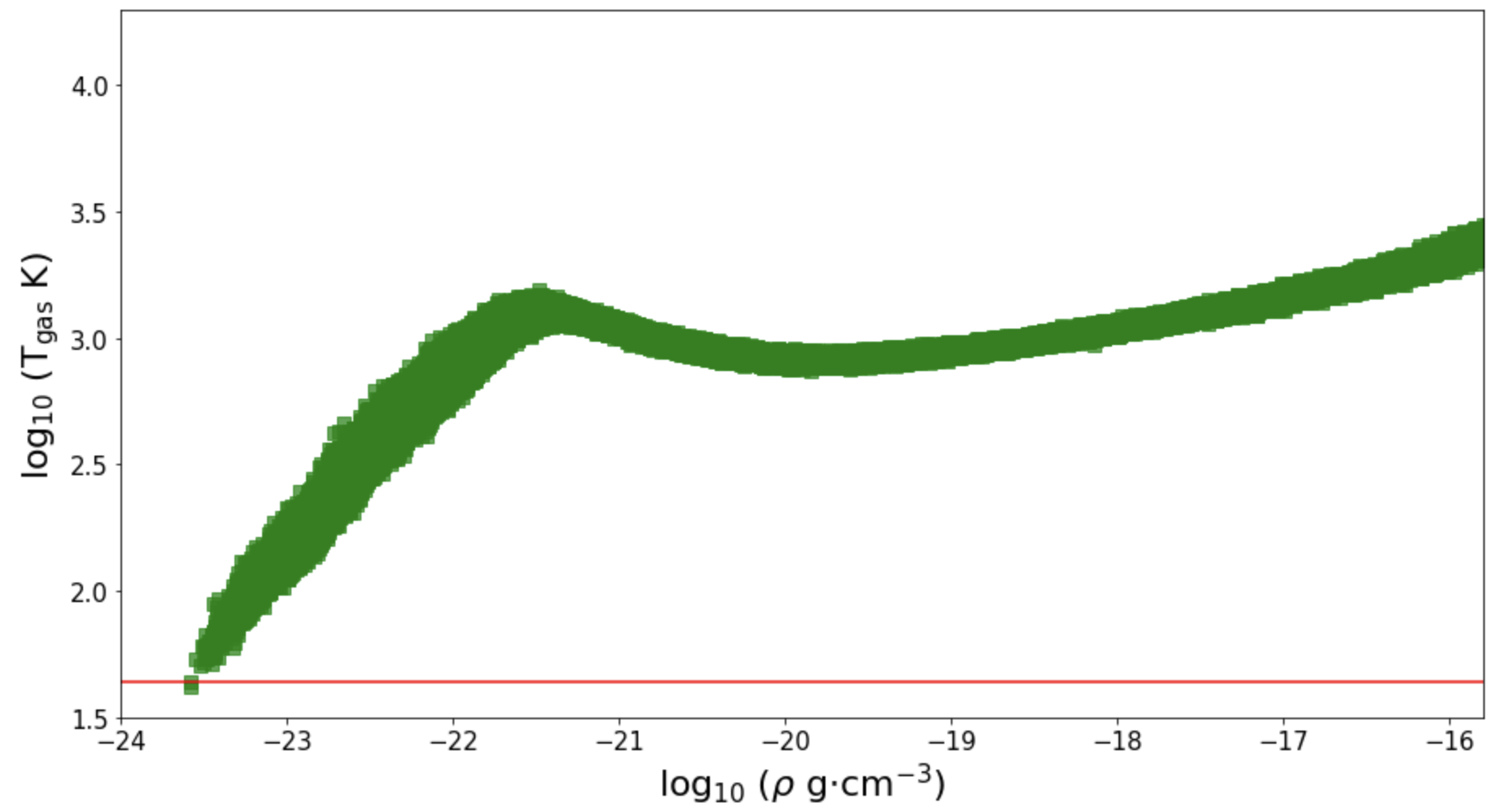}
\caption{Density profiles of temperature evolution of the prestellar cloud of the primordial model resulting from the hydrodynamical runs. The red bottom line represents the CMB floor temperature give by the initial redshift of the simulations.}
\label{hydro_rho_T}
\end{figure}

Fig. \ref{n_T} shows the density profile of the temperature evolution for different cloud models resulting from the one-zone simulations. The dotted line represents the primordial model, the dotted-dashed line the met3 model, the dashed line the met2 model, and the solid line the Keller model. The red bottom line represents the CMB floor temperature given by the initial redshift of the simulations. The temperature is given in K, while the number density in cm$^{-3}$. From the figure, the enhancement in the cooling rate for the cloud is evident, as even for a slight presence of metals as $Z/Z_{\odot}=10^{-3}$ the temperature of the cloud drops drastically compared to the primordial model at densities of $\sim 10^{3} \mathrm{cm}^{-3}$. Moreover, for a metallicity $Z/Z_{\odot}=10^{-2}$ the cloud is already able to reach the CMB floor temperature, in agreement with previous results. Further, the temperature evolution of the Keller model is very similar to the met2 model, due to the high presence of carbon. Fig. \ref{hydro_rho_T} shows the density profile of the termperature's evolution for the primordial model resulting from the hydrodynamical runs. The red solid bottom line represents the CMB floor temperature.

\section{Discussion}
\label{ch4}
We have presented the coupling between KROME and GRADSPH (Fig. \ref{hydro_rho_T}), as well as results using only the former (Fig. \ref{n_T}). By looking at Fig. \ref{n_T} the enhancement in the cooling of the clouds it is evident, which is consistent with previous results (e.g., \cite{Bovino2014}). Further, the results from the one-zone runs are in agreement with the metallicity threshold proposed by \cite{Bromm2001}, showing that for a metallicity $Z/Z_{\odot}=10^{-2}$, clouds are already able to reach the CMB floor temperature. In addition, the high presence of ionized carbon and neutral oxygen in the Keller model allows to the cloud to follow closely the temperature evolution of the met2 model. This reflects their major contribution as cooling channels at high redshift star-forming conditions, which is in agreement with previous results.\\

In order to improve the accuracy of the results generated by the simulation it is necessary to include further physical processes such as the presence of a UV radiation background or the treatment of a dust grain distribution. The former has been proven to have a lesser impact on the thermodynamics of the collapse of a cloud, but this need to be confirmed by further studies. The latter has been shown to have major impact of the thermodynamics of a collapsing cloud by several authors, showing its great impact as a catalyzer for several chemical reactions that are impossible without the presence of a third-body, or acting as a shield for the external radiation that hits the cloud.

%%%%%%%%%%%%%%%%%%%%%%%%%%%%%%%%%%%%%%%%%%%%%%%%%%%%%%%%%%%%%%%%%%%%%%%%%%%%%%
%                                                                            %
% Para figuras de dos columnas use \begin{figure*} ... \end{figure*}         %
%                                                                            %
%%%%%%%%%%%%%%%%%%%%%%%%%%%%%%%%%%%%%%%%%%%%%%%%%%%%%%%%%%%%%%%%%%%%%%%%%%%%%%

\begin{acknowledgement}
The simulations were performed with resources provided by the Kultrun Astronomy Hybrid Cluster via the projects Conicyt Programa de Astronomia Fondo Quimal 2017 (project code QUIMAL170001), Conicyt PIA (project code ACT172033), and Fondecyt Iniciacion (project code 11170268). Powered@NLHPC: This research was partially supported by the supercomputing infrastructure of the NLHPC (ECM-02). DRGS, SB, VD, CO and FF thank for funding via CONICYT PIA ACT172033. DRGS and SB acknowledge funding through CONICYT project Basal AFB-170002. RR, CO, FF and DRGS thank for funding through the 'Concurso Proyectos Internacionales de Investigaci\'on, Convocatoria 2015' (project code PII20150171). DRGS, SB and VBD thank for funding via CONICYT PIA ACT172033. FF and VBD thank for funding through Fondecyt regular (project code 1161247). VBD thanks to Conicyt for financial support on her master studies (CONICYT-PFECHA/Mag\'isterNacional/2017-22171293). 
\end{acknowledgement}

%%%%%%%%%%%%%%%%%%%%%%%%%%%%%%%%%%%%%%%%%%%%%%%%%%%%%%%%%%%%%%%%%%%%%%%%%%%%%%
%                                                                            %
%  Por favor no modifique las líneas de la bibliografía, salvo el nombre     %
%  el archivo de Bibtex con la lista de citas (sin la extensión .BIB)        %
%                                                                            %
%  Please do not modify the following lines, except the name of the Bibtex   %
%  file (whithout the .BIB extension)                                        %
%                                                                            %
%%%%%%%%%%%%%%%%%%%%%%%%%%%%%%%%%%%%%%%%%%%%%%%%%%%%%%%%%%%%%%%%%%%%%%%%%%%%%% 

\bibliographystyle{baaa}
\small
\bibliography{biblio}
 
\end{document}